\documentclass{article}[12]                                                 
\begin{document}     
\begin{center}{\Large\bf Cosmological Relativity: Determining the Universe
by the Cosmological Redshift}\end{center}
\begin{center}{Moshe Carmeli}\end{center}
\begin{center}{Department of Physics, Ben Gurion University, Beer Sheva 84105, 
Israel}\end{center}
\begin{center}{Email: carmelim@bgumail.bgu.ac.il}\end{center}
\begin{abstract}
Using cosmological relativity theory, we derive the formula for the 
cosmological redshift written explicitly in terms of $(1-\Omega)$, where 
$\Omega=\rho/\rho_c$ is the ratio of the average mass density to the critical
``closure" density. Based on the present-day data of observed redshifts, we
conclude that $\Omega<1$.
\end{abstract}
\section{Introduction}
In spite of the advances made in recent years in cosmology, the question of
what kind of universe we live in is still unsettled [1-10]. According to FRW, based 
on general relativity theory, one can calculate a critical mass density 
$\rho_c$ at the nowadays time such that the Hubble expansion will ultimately
be reversed if, and only if, the actual average mass density $\rho$ exceeds
$\rho_c$, where $\rho_c=3H_0^2/8\pi G$, with $H_0$ the present-time Hubble
parameter and $G$ the Newtonian gravitational constant. The value of $\rho_c$
is about $10^{-29}$g/cm$^3$, a few hydrogen atoms per cubic meter throughout
the cosmos. Thus it is the value of $\Omega=\rho/\rho_c$ which determines the
behavior of the expansion of the universe: $\Omega>1$, a finite universe,
$\Omega<1$, an infinite curved universe, $\Omega=1$ an infinite flat-space
universe, and the sign of the quantity $(1-\Omega)$ is the determining factor
here [11,12].

In this paper we use cosmological general relativity theory [13-15] to derive
a general formula for the redshift in which the term $(1-\Omega)$ appears 
explicitly. Since there are enough data of measurements of redshifts, this 
allows one to determine what is the sign of $(1-\Omega)$, positive, zero or 
negative. Our conclusion is that $(1-\Omega)$ cannot be negative or zero. This
means that the universe is infinite, curved and expands forever, a result
favored by some cosmologists [16]. To this end we proceed as follows.
\section{Gravitational Field Equations}
We seek a spherical symmetric solution to the Einstein field equations 
$G_\mu^\nu=\kappa T_\mu^\nu$ and use spherical coordinates $x^\mu=(x^0,x^1,
x^2,x^3)=(\tau v,r,\theta,\phi)$, where $\tau$ is Hubble's time in the 
zero-gravity limit and $v$ is the velocity parameter. Since the universe is spherically symmetric at
any chosen point, the line element we seek is of the form
$$ds^2=\tau^2dv^2-e^\lambda dr^2-r^2\left(d\theta^2+\sin^2\theta d\phi^2
\right),\eqno(1)$$
where comoving coordinates, as in the Friedmann theory, are used and $\lambda$
is a function of the radial distance $r$ only. To determine $\lambda$ it is enough to solve the 
field equation $G_0^0=\kappa T_0^0$ which turns out to be [13-15]
$$G_0^0=e^{-\lambda}\left(\frac{\lambda'}{r}-\frac{1}{r^2}\right)+
\frac{1}{r^2}=\frac{8\pi G}{c^4}T_0^0=\frac{8\pi G}{c^2}\rho_{eff},\eqno(2)$$
where a prime denotes derivation with respect to $r$ and $\rho_{eff}=\rho-
\rho_c$. The solution of Eq. (2) is
$$e^{-\lambda}=1+\frac{\left(1-\Omega\right)}{c^2\tau^2}r^2,\eqno(3)$$
with $g_{11}=-e^{\lambda}$, $a^2=c^2\tau^2/(1-\Omega)$ and $\Omega=\rho/\rho_c$. 
\section{Cosmological Redshift}
Having the metric tensor we may now find the redshift of light emitted in the
cosmos. As usual, at two points 1 and 2 we have for the wave lengths:
$$\frac{\lambda_2}{\lambda_1}=\frac{ds\left(1\right)}{ds\left(2\right)}=
\sqrt{\frac{g_{11}\left(1\right)}{g_{11}\left(2\right)}}.\eqno(4)$$
Using now the solution for $g_{11}$ in Eq. (4) we obtain
$$\frac{\lambda_2}{\lambda_1}=\sqrt{\frac{1+r_2^2/a^2}{1+r_1^2/a^2}}.\eqno(5)$$

For a sun-like body located at the coordinates origin, and an observer at a
distance $r$ from the center of the body, we then have $r_2=r$ and $r_1=0$, 
thus 
$$\frac{\lambda_2}{\lambda_1}=\sqrt{1+\frac{r^2}{a^2}}=
\sqrt{1+\frac{\left(1-\Omega\right)r^2}{c^2\tau^2}} \eqno(6)$$
for the cosmological contribution to the redshift. If, furthermore, $r\ll a$
we then have
$$\frac{\lambda_2}{\lambda_1}=1+\frac{r^2}{2a^2}=
1+\frac{\left(1-\Omega\right)r^2}{2c^2\tau^2},\eqno(7)$$
to the lowest apprixomation in $r^2/a^2$, and thus 
$$z=\frac{\lambda_2}{\lambda_1}-1=\frac{r^2}{2a^2}=
\frac{\left(1-\Omega\right)r^2}{2c^2\tau^2}.\eqno(8)$$

From Eqs. (6)--(8) it is clear that $\Omega$ cannot be larger than one since
otherwise $z$ will be negative, which means blueshift, and as is well known
nobody sees such a thing. If $\Omega=1$ then $z=0$, and for $\Omega<1$ we have 
$z>0$. The case of $\Omega=1$ is also implausible since the light from stars
we see is usually redshifted more than the redshift due to the gravity of the 
body emitting the radiation, as is evident from our sun, for example, whose
emitted light is shifted by only $z=2.12\times 10^{-16}$ [17].
\section{Conclusions}
One can thus conclude that the theory of cosmological  general relativity
predicts that the universe  is infinite and expands from now on 
forever. As is well known the standard FRW model does not relate the 
cosmological redshift to the kind of the universe. Our conclusion is also in
full agreement with the measurements recently obtained by the {\it High-Z
Supernovae Team} and the {\it Suprenovae Cosmology Project} [18-24].

\end{document}